# Large, Homogeneous, and Isotropic Critical Current Density in Oxygen-Annealed $Fe_{1+y}Te_{0.6}Se_{0.4}$ Single Crystal

Yue Sun[1,2], Toshihiro Taen[2], Yuji Tsuchiya[2], Qingping Ding[2], Sunsen Pyon[2], Zhixiang Shi[1*], and Tsuyoshi Tamegai[2*]

[1]*Department of Physics, Southeast University, Nanjing 211189, People's Republic of China*

[2]*Department of Applied Physics, The University of Tokyo, 7-3-1 Hongo, Bunkyo-ku, Tokyo 113-8656, Japan*

[*]E-mail addresses: zxshi@seu.edu.cn and tamegai@ap.t.u-tokyo.ac.jp

## *Abstract*

We reported a controllable way of removing excess Fe in $Fe_{1+y}Te_{0.6}Se_{0.4}$ by annealing in a fixed amount of $O_2$. Compared with the weak superconductivity induced by dilute acids, $O_2$ annealing can successfully induce bulk superconductivity. Proper $O_2$ annealing changes the temperature dependence of resistivity at low temperatures from semiconducting to metallic, which comes from the deintercalation of excess Fe. Critical current densities with field along the *c*-axis and *ab*-plane are found almost isotropic with large values of $J_c^{H//c} \sim 3 \times 10^5$ A/cm$^2$ and $J_c^{H//ab} \sim 2.5 \times 10^5$ A/cm$^2$ at 2 K. Furthermore, magneto-optical images reveal isotropic current flow within the *ab*-plane.

Stimulated by the discovery of high-temperature superconductivity in LaFeAsO$_{1-x}$F$_x$,[1] a series of iron-based superconductors have been reported with four main structures: 1111 (LaFeAs(O,F)), 122 ((Ba,K)Fe$_2$As$_2$), 111 (LiFeAs) and 11 (Fe(Te,Se)).[2] Among these, FeTe$_{1-x}$Se$_x$ composed of only Fe(Te,Se) layers has attracted special attention due to its simple crystal structure, which is preferable for probing the mechanism of superconductivity. On the other hand, its less toxic nature makes FeTe$_{1-x}$Se$_x$ a more suitable candidate for applications among the family of iron-based superconductors. Although much attention has been paid to this compound, some crucially fundamental physical properties as well as the phase diagram are still controversial. Although Liu et al.[3] reported that bulk superconductivity resides only in the region of Se level higher than 29%, it was observed in the Se doping region between 10 and 50% by Noji et al.[4, 5] As for the Hall coefficient, conflicting low-temperature behaviors were observed in crystals with nominally the same amount of Se.[6-8] Even in the case of resistivity, both the metallic and nonmetallic behaviors were observed, and the absolute value just above $T_c$ has a spread from 200 to 1500 $\mu\Omega$ cm.[5-10] These controversies come from the sample-dependent Fe nonstoichiometries,[11, 12] which originate from the partial occupation of excess Fe at the interstitial site in the Te/Se layer. In FeTe$_{1-x}$Se$_x$, the magnetic coupling between the excess Fe and the adjacent Fe sheets will not only suppress the superconductivity but also localize the charge carriers.[7] The effect of Fe nonstoichiometry may also be responsible for the relatively small critical current density in FeTe$_{1-x}$Se$_x$ bulks and wires,[13, 14] which is much smaller than that in Ba$_{0.6}$K$_{0.4}$Fe$_2$As$_2$ wires (~ 1.2 × 10$^5$ A/cm$^2$).[15] Furthermore, no homogeneous distribution of critical current density has been observed yet even in single crystals, which casts a shadow over the application research of iron chalcogenides.

To remove the effect of Fe nonstoichiometry, some attempts have been made by annealing crystals under different conditions.[5, 10, 16-18] Although superconductivity can be induced by several methods, some annealed samples show a very broad transition width, which implies an inhomogeneous superconductivity. In a previous work, we found that annealing in O$_2$ atmosphere is very effective to induce bulk superconductivity in Fe$_{1+y}$Te$_{0.6}$Se$_{0.4}$.[19] In this paper, we reported a controllable way of removing the excess Fe by annealing the crystal with a fixed amount of O$_2$. Some properties under controversy were recharacterized in the well-annealed crystal. A large, homogeneous, and almost isotropic critical current density was revealed for the first time in iron chalcogenides.

Single crystal with a nominal composition FeTe$_{0.6}$Se$_{0.4}$ was grown by the self-flux method as reported previously.[19] For O$_2$ annealing, the sample was loaded into a quartz tube, which was carefully baked and examined that no appreciable amount of gas was emitted under the same condition as the sample annealing. The quartz tube was carefully evacuated using a diffusion pump and filled with a fixed amount of O$_2$ gas before sealing the quartz tube. By measuring the volume of the quartz tube and the weight of the crystal, the molar ratio of oxygen to the total amount of Fe was determined to be ~ 1.5 %. During these processes, a diaphragm-type manometer with an accuracy higher than 1 mTorr was used for real-time monitoring the pressure in the system to prevent gas leakage and to check the amount of O$_2$ in the tube. Then, the samples were annealed at 400 ℃ for more than one day, followed by water quenching. Other pieces of as-grown samples were put into glass bottles (10 ml) filled with either 20 % hydrochloric acid (HCl) or 5 % nitric acid (HNO$_3$). The sample immersed in 20% HCl was kept at room temperature for 100 h. On the other hand, the sample immersed in 5 % HNO$_3$ was kept at room temperature for just 30 mins, because we found that the sample itself was damaged when it was immersed in HNO$_3$ for a long time. Magnetization measurements on the as-grown and post-treated samples were performed using a commercial superconducting quantum interference device (SQUID) magnetometer (MPMS-XL5, Quantum Design). Resistivity measurements were performed by the four-probe method. Magneto-optical (MO) images were obtained by using the local field-dependent Faraday effect in the in-plane magnetized garnet indicator film employing a differential method. Inductively-coupled plasma (ICP) atomic emission spectroscopy and energy dispersive X-ray spectroscopy (EDX) were used for chemical analyses.

The inset of Fig. 1 shows the temperature dependences of zero-field-cooled (ZFC) and field-cooled (FC) magnetizations at 5 Oe for the as-grown, HNO$_3$- and HCl-treated, and O$_2$-annealed Fe$_{1+y}$Te$_{0.6}$Se$_{0.4}$ single crystals. The as-grown crystals usually show no superconductivity or a very weak diamagnetic signal only below 3 K.[10, 19] Although superconductivity can be successfully induced by immersing the sample in HNO$_3$ or HCl, $T_c$ is lower and the transition width is much broader than that in the crystal annealed in a fixed amount of O$_2$, which shows a $T_c$ higher than 14 K with the transition width less than 1 K (obtained from the criteria of 10 and 90% of the magnetization). Here, we must point out that when the sample was annealed in too much oxygen, it was completely oxidized and turned into an insulator.

To further confirm the nature of superconductivity, critical current density, $J_c$ [A/cm$^2$], was obtained from magnetic hysteresis loops (MHLs) by using the extended Bean model:

$$J_c = 20\frac{\Delta M}{a(1-a/3b)}, \qquad (1)$$

where $\Delta M$ is $M_{\text{down}} - M_{\text{up}}$, $M_{\text{up}}$ [emu/cm$^3$] and $M_{\text{down}}$ [emu/cm$^3$] are the magnetizations when sweeping fields up and down, respectively, and $a$ [cm] and $b$ [cm] are sample widths ($a < b$). The magnetic field dependence of $J_c$ at 2 K is shown in the main panel of Fig. 1. The $J_c$ of the O$_2$-annealed sample reaches about $3\times 10^5$ A/cm$^2$ at zero field. The magnetic field dependence of $J_c$ is weak, maintaining a value of $\sim 1\times 10^5$ A/cm$^2$ even at 50 kOe. Although the $J_c$ of O$_2$-annealed Fe$_{1+y}$Te$_{0.6}$Se$_{0.4}$ single crystal is lower than that of the Ba(Fe$_{1-x}$Co$_x$)$_2$As$_2$ single crystal,[20] it is one of the largest values among those reported in Fe(Te,Se).[19] It demonstrates the high quality of our O$_2$-annealed sample. On the other hand, $J_c$ in crystals immersed in HNO$_3$ and HCl at 2 K under zero field is smaller than $1\times 10^4$ A/cm$^2$. In addition, $J_c$ decreases very quickly with increasing field and becomes smaller than $1\times 10^2$ A/cm$^2$ at fields above 15 kOe. The much smaller value of $J_c$ than that of the O$_2$-annealed sample indicates that acids can just induce superconductivity near the surface layers similarly to the case of alcoholic beverages and I$_2$ atmosphere.[19] We speculated that their relatively larger size prevents them from intercalating between the layers of the crystal. Thus, they can just oxidize the excess Fe close to the surface or edge of the crystal.[19]

Figure 2(a) shows the temperature dependence of in-plane resistivity for the as-grown and O$_2$-annealed Fe$_{1+y}$Te$_{0.6}$Se$_{0.4}$ single crystals. For the as-grown sample, the resistivity maintains a nearly constant value of $\sim$ 600 µΩ cm at higher temperatures, followed by a nonmetallic behavior ($d\rho/dT < 0$) below 120 K down to the superconducting transition. The upturn of resistivity above $T_c$ is usually observed in Fe$_{1+y}$Te$_{1-x}$Se$_x$ single crystals or films with relatively lower $T_c$'s,[6, 8, 21] which may originate from the localization of the charge carriers by disorders induced by the magnetic moment of excess Fe.[6] In this work, we found that the nonmetallic resistivity behavior above $T_c$ is suppressed by O$_2$ annealing and replaced by a metallic behavior ($d\rho/dT > 0$) below 200 K. The absolute value of the resistivity is also decreased by O$_2$ annealing. This may be caused by the delocalization of charge carriers by removing excess Fe from the crystal.[19] The resistivity at $T_c$ is $\sim$ 180 µΩ cm, which is the smallest among those reported in Fe(Te,Se) crystals and films.[3, 6, 8-10, 22] Although this value is still slightly larger than that in typical Ba(Fe$_{1-x}$Co$_x$)$_2$As$_2$ single crystals and films (~ 100 μΩ cm),[23] it is very close to that in Ba(Fe$_{1-x}$Co$_x$)$_2$As$_2$ with disorder in FeAs planes.[24] The smallest resistivity just above $T_c$ indicates that these crystals have more perfect conducting Fe(Te,Se) layers. Based on the report by Liu *et al.* that the as-grown Fe$_{1+y}$Te$_{0.6}$Se$_{0.4}$ single crystal with less excess Fe shows metallic behavior, while that with more excess Fe shows nonmetallic behavior,[6] we judge that the O$_2$ annealing process removes excess Fe from the bulk of the sample. However, we believe that the excess Fe still remains in the sample, since ICP analyses on the O$_2$ annealed crystal show no change in the Fe content. In addition, after controlled O$_2$ annealing, we found the surface layers of the crystal change color to dark blue. A similar color change has been reported in the sample annealed in the air.[25] Furthermore, our EDX results show the blue surface layers have more excess Fe than the inner part, which may indicate that iron oxide formation on the surface may be responsible for the color change. Actually, oxygen was evidenced by electron energy loss spectroscopy in the O$_2$-annealed Fe$_{1.08}$Te$_{0.55}$Se$_{0.45}$ crystal by Hu *et al.*[26]

The temperature dependences of resistivity with applied field from 0 to 50 kOe for $H \parallel ab$ and $H \parallel c$ are shown in the upper and lower insets of Fig. 2(b), respectively. For $H \parallel c$, with increasing field, the resistive transition shifts to lower temperatures accompanied by a slight increase in the transition width, while this broadening is almost negligible for $H \parallel ab$. The upper critical field $H_{c2}$ for $H \parallel ab$ and $H \parallel c$ defined by the midpoint of resistive transition is plotted in Fig. 2(b). The slopes of $H_{c2}$ are -55 and -99 kOe/K along the *c*-axis and *ab*-plane, respectively, extracted from the linear part above 20 kOe. The anisotropy parameter γ close to the $T_c$ is estimated as 1.8, which is close to that reported before.[5, 27] According to the Werthamer-Helfand-Hohenberg theory,[28] $H_{c2}$ limited by the orbital depairing in the dirty limit is given by $H_{c2}(0) = -0.693 T_c dH_{c2}/dT \big|_{T=T_c}$. The $H_{c2}^c(0)$ and $H_{c2}^{ab}(0)$ are estimated as ~565 and ~1020 kOe, respectively using extrapolated $T_c$ from the linear part. The large upper critical fields and small anisotropy show that Fe(Te,Se) is a suitable candidate for applications.

To further probe the superconducting and magnetic properties of the O$_2$-annealed Fe$_{1+y}$Te$_{0.6}$Se$_{0.4}$ single crystal, magnetic field dependent $J_c$'s for $H \parallel c$ and $H \parallel ab$ obtained from MHLs are shown in Figs. 3(a) and 3(b), respectively. Actually, in tetragonal two-dimensional systems, there are three kinds of critical current, $J_c^{x,y}$, where $x$ and $y$ refer to the directions of current and magnetic field, respectively. For $H // c$, irreversible magnetization is determined solely by $J_c^{ab,c}$. This means that $J_c^{ab,c}$ ($= J_c^{H//c}$) can be easily evaluated from the measured magnetization using eq. (1). On the other hand, in the case of $H // ab$, both $J_c^{ab,ab}$ and $J_c^{c,ab}$ contribute to the measured magnetization. Here, we assume that $J_c^{ab,ab}$ is equal to $J_c^{c,ab}$ and obtain the weighted average $J_c^{H//ab}$ using eq. (1). Based on the discussion above, self-field critical current densities $J_c^{H//c}$ and $J_c^{H//ab}$ at 2 K are estimated as $3 \times 10^5$ and $2.5 \times 10^5$ A/cm$^2$, respectively. Thus, the critical current densities are almost isotropic for the O$_2$-annealed Fe$_{1+y}$Te$_{0.6}$Se$_{0.4}$ single crystal at low temperatures and low fields. All these behaviors of $J_c$ in the O$_2$-annealed Fe$_{1+y}$Te$_{0.6}$Se$_{0.4}$ are similar to that in Ba(Fe$_{0.93}$Co$_{0.07}$)$_2$As$_2$.[29]

The above estimations of $J_c$'s using the Bean model rely on the assumption that homogeneous currents flow within the sample. To examine this assumption, we took MO images on the same piece of single crystal in the remanent state at temperatures ranging from 5 to 14 K. This state is prepared by applying 400 Oe along the c-axis for 1 s and removing it after zero-field cooling. Typical MO images from 5 to 14 K are shown in Figs. 4(a) – 4(d). At 5 K, the field cannot penetrate the sample to the center. As the temperature is increased, the field penetrates deeper into the center. At 8 K, the sample is totally penetrated, and the MO image manifests a typical roof-top pattern up to 10 K, similar to that observed in high-quality Ba(Fe$_{1-x}$Co$_x$)$_2$As$_2$ single crystals,[24, 29] indicating a nearly uniform current flow in the crystal. Besides, the typical current discontinuity lines (so-called d-line), which cannot be crossed by vortices, can be observed and marked by the dotted line in Fig. 4 (b). By measuring the angles of the discontinuity line for our rectangular sample, the in-plane anisotropy of the current densities can be easily estimated.[30] In the present case, the angle $\theta$ is ~ 45°, indicating that the critical current density within the ab-plane is isotropic, consistent with the fourfold symmetry of the superconducting plane. As temperature approaches $T_c$, although the roof-top pattern cannot be seen clearly, some parts of the crystal can still trap field as shown in Fig. 4(d) for the image taken at 14 K. Such large value and homogeneous distribution of critical current density are observed in iron chalcogenide compounds for the first time, which again proves the high quality of our O$_2$-annealed Fe$_{1+y}$Te$_{0.6}$Se$_{0.4}$ single crystal and guarantees the potential application of this compound. Figure 4 (e) shows profiles of the magnetic induction at different temperatures along the dashed line shown in Fig. 4(a). In the case where the magnetic field does not penetrate into the center of the sample, the critical current density can be roughly estimated using the following equation for thin superconducting strips,

$$J_c = \frac{c}{4d} \frac{H_{ex}}{\cosh^{-1}[w/(w-2p)]}, \quad (2)$$

where $H_{ex}$ is the external field, $w$ is the sample width, $d$ is the sample thickness, and $p$ is the penetration distance measured from the sample edge.[31] For the case where the magnetic field penetrates into the whole sample, $J_c$ can be roughly estimated by $J_c \sim \Delta B/d$, where $\Delta B$ is the trapped field and $d$ is the thickness of the sample. $J_c$ obtained from the MO images is consistent with that estimated from MHLs, as shown in Fig. 4(f).

In summary, we found a controllable way of removing the excess Fe in $Fe_{1+y}Te_{0.6}Se_{0.4}$ single crystal by annealing with a fixed amount of $O_2$. Compared with the weak superconductivity induced by dilute acids, $O_2$ annealing is proved to be effective for inducing bulk superconductivity. The $O_2$ annealing changes the nonmetallic resistive behavior into a metallic one with a reasonably low resistivity of ~ 180 μΩ cm at $T_c$. Critical current densities with field along the *c*-axis and *ab*-plane are almost isotropic with large values of $J_c^{H//c} \sim 3 \times 10^5$ A/cm$^2$ and $J_c^{H//ab} \sim 2.5 \times 10^5$ A/cm$^2$ at 2 K. Magneto-optical images reveal homogenous and isotropic current flow within the *ab*-plane.

**Acknowledgment** This work was partly supported by the Natural Science Foundation of China, the Ministry of Science and Technology of China (973 project: No. 2011CBA00105), and the Japan-China Bilateral Joint Research Project by the Japan Society for the Promotion of Science.

# Figure captions

Fig. 1: Field dependence of $J_c$ at 2 K for $Fe_{1+y}Te_{0.6}Se_{0.4}$ immersed in $HNO_3$ and HCl, as well as annealed in $O_2$. The inset shows the temperature dependences of zero-field-cooled (ZFC) and field-cooled (FC) magnetizations at 5 Oe for the three post-treated samples together with the as-grown one.

Fig. 2: (a) Temperature dependence of in-plane resistivity in the as-grown and $O_2$-annealed $Fe_{1+y}Te_{0.6}Se_{0.4}$. (b) Temperature dependence of upper critical fields along *ab* and *c* directions. Upper and lower insets show that magnetic field dependences of in-plane resistivity for *H // ab* and *H // c*, respectively.

Fig. 3: Magnetic field dependences of critical current densities of $O_2$-annealed $Fe_{1+y}Te_{0.6}Se_{0.4}$ for (a) *H // c* and (b) *H // ab*.

Fig. 4: Magneto-optical images in the remanent state after applying 400 Oe along the *c*-axis in $O_2$-annealed $Fe_{1+y}Te_{0.6}Se_{0.4}$ at (a) 5, (b) 8, (c) 10, and (d) 14 K. (e) Local magnetic induction profiles at temperatures from 5 to 14 K taken along the dashed lines in (a). (f) Temperature dependence of $J_c$ estimated from MHLs and MO measurements.

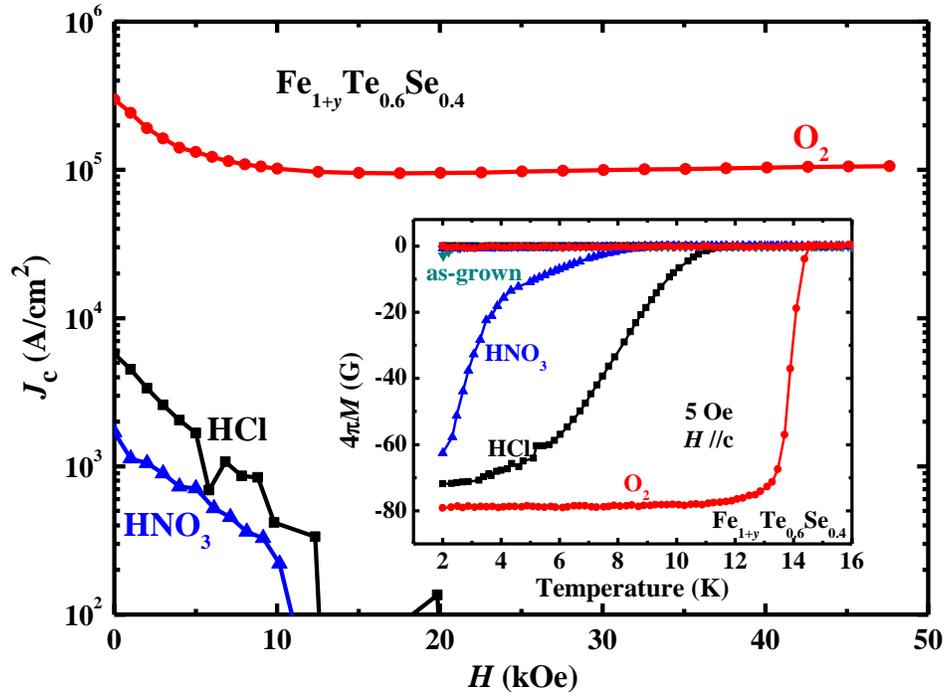

**Figure. 1 Y. Sun *et al.***
11

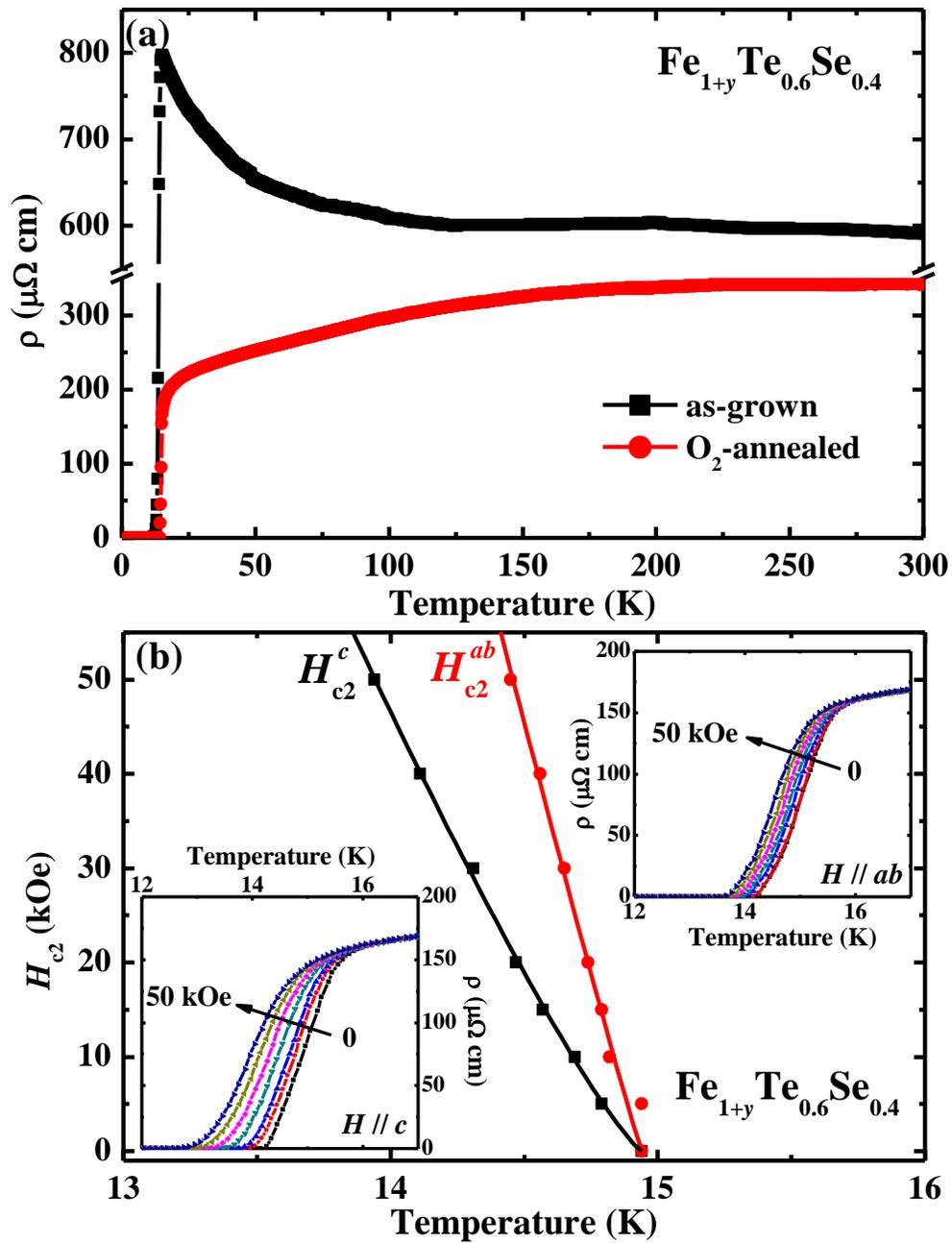

**Figure. 2** Y. Sun *et al.*

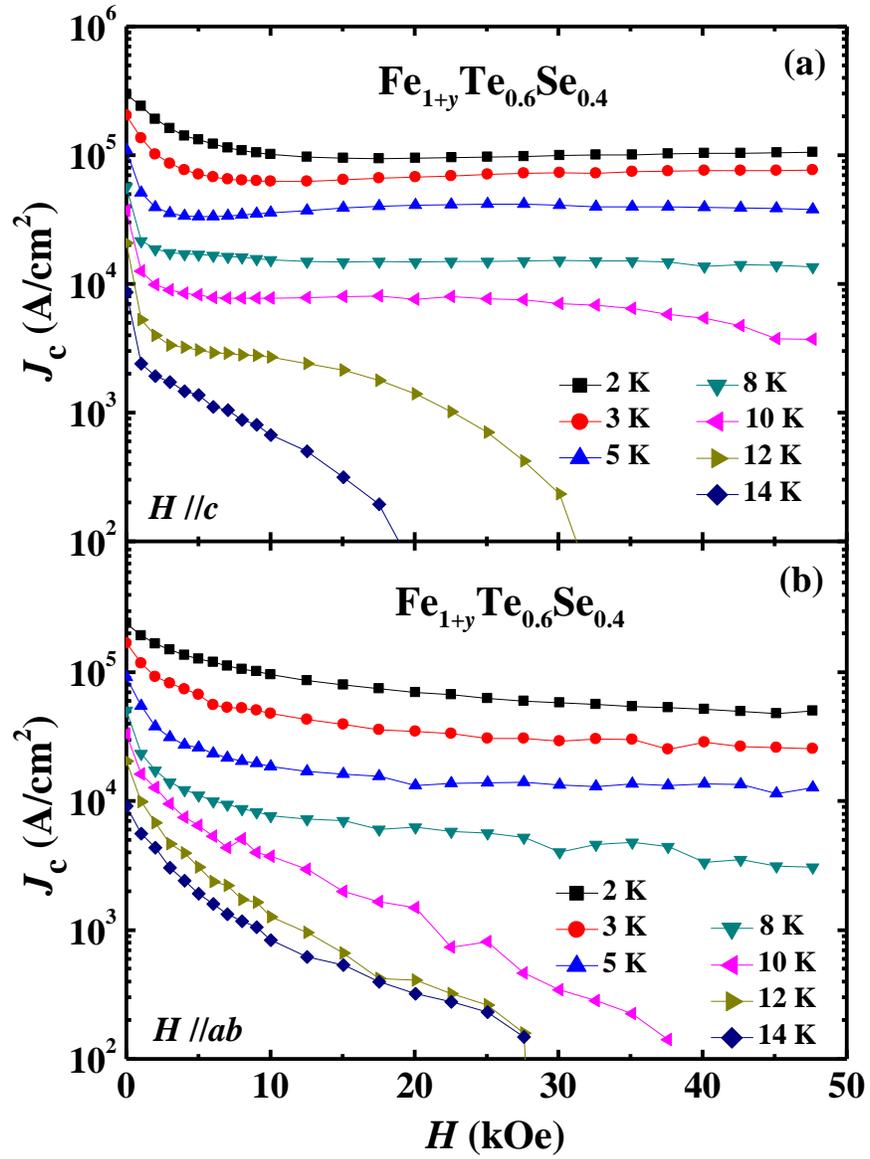

Figure. 3 Y. Sun *et al.*

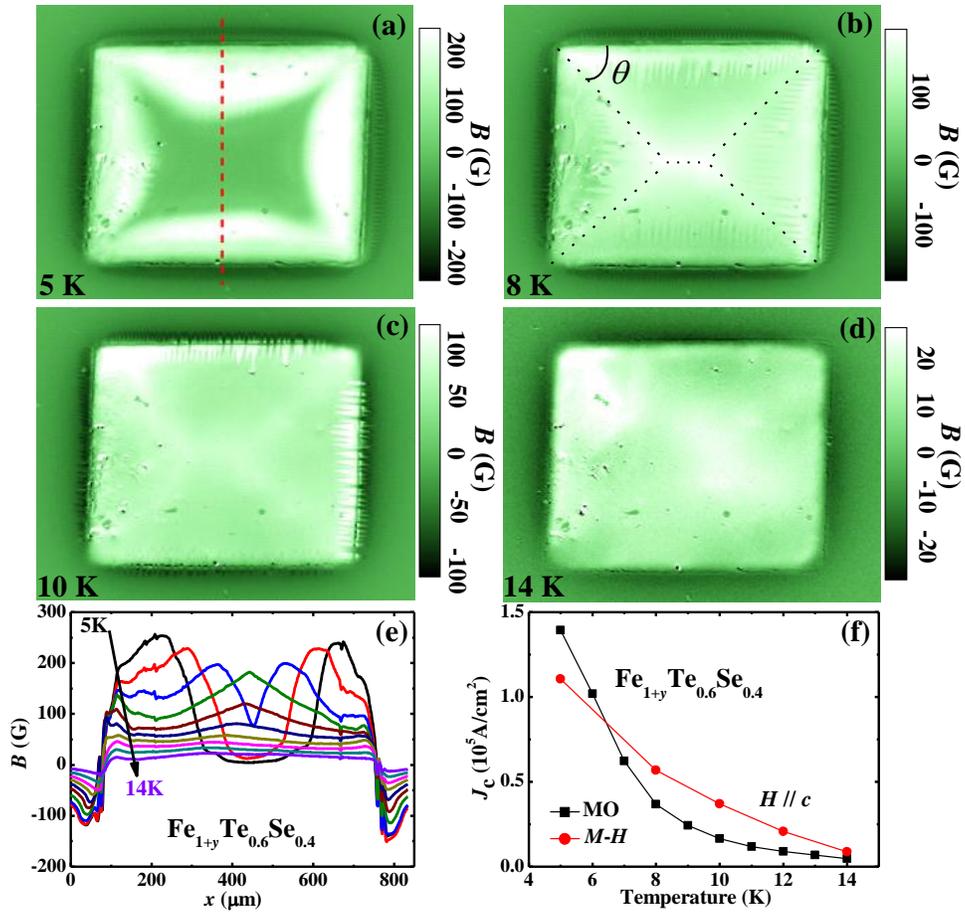

**Figure. 4 Y. Sun** *et al.*